\documentclass[prl,twocolumn,showpacs,preprintnumbers,amsmath,amssymb]{revtex4}

\usepackage{graphicx}
\usepackage{dcolumn}
\usepackage{bm}
\usepackage{verbatim}
\usepackage{color}
\usepackage{hyperref}

\newcommand{\dd}{{\rm d}}

\newcommand{\cT}{{\mathcal{T}}}
\newcommand{\cD}{{\mathcal{D}}}
\newcommand{\cS}{{\mathcal{S}}}

\newcommand{\bR}{{\mathbb{R}}}

\newcommand{\mean}[1]{\overline{#1}}

\newcommand{\beq}{\begin{equation}}
\newcommand{\eeq}{\end{equation}}

\newcommand{\beqS}{\begin{equation*}}
\newcommand{\eeqS}{\end{equation*}}

\newcommand{\beqa}{\begin{eqnarray}}
\newcommand{\eeqa}{\end{eqnarray}}

\newcommand{\beqaS}{\begin{eqnarray*}}
\newcommand{\eeqaS}{\end{eqnarray*}}

\usepackage{amsmath}

\begin{document}

\title{Railway switch transport model}

\author{Martin Horvat}
\email{martin.horvat@fmf.uni-lj.si}

\author{Toma\v z Prosen}
\email{tomaz.prosen@fmf.uni-lj.si}
\affiliation{Faculty of mathematics and physics, Department of Physics, University of Ljubljana,  Ljubljana, Slovenia}

\author{Giuliano Benenti }
\email{giuliano.benenti@uninsubria.it}
\affiliation{CNISM and Center for Nonlinear and Complex Systems,
Universit\`a degli Studi dell'Insubria, via Valleggio 11, 22100 Como, Italy}
\affiliation{Istituto Nazionale di Fisica Nucleare, Sezione di Milano,
via Celoria 16, 20133 Milano, Italy}

\author{Giulio Casati}
\email{giulio.casati@uninsubria.it}
\affiliation{CNISM and Center for Nonlinear and Complex Systems,
Universit\`a degli Studi dell'Insubria, via Valleggio 11, 22100 Como, Italy}
\affiliation{Istituto Nazionale di Fisica Nucleare, Sezione di Milano,
via Celoria 16, 20133 Milano, Italy}

\date{\today}

\begin{abstract}
We propose a simple model of coupled heat and particle transport based on 
a zero-dimensional classical deterministic dynamics which is reminiscent 
of a railway switch whose action is only a function of the particle's energy.  
It is shown that already in the minimal three-terminal model, where the second 
terminal is considered as a probe with zero net particle and heat currents, 
one can find extremely asymmetric Onsager matrices as a consequence of 
time-reversal symmetry breaking of the model. This minimalistic transport 
model provides a better understanding of thermoelectric 
heat engines in the presence of time-reversal symmetry breaking.
\end{abstract}



\maketitle

{\em Introduction.-}  Minimalistic mathematical models often provide key 
paradigms in theoretical physics. In the theory of coherent quantum transport 
\cite{datta} of non-interacting electron systems, the conductances can be 
elegantly formulated solely in terms of transmission and reflection 
amplitudes \cite{buttiker}.
Very similar expressions for conductances can be written also in the realm of 
classical physics, if the particle's dynamics inside the systems is 
deterministic and conservative (Hamiltonian) (see e.g. \cite{horvat}). 
In multi-terminal transport theory, probe terminals \cite{buttiker_probe} 
are conveniently used as minimalistic models of inelastic scattering. 
A probe is a terminal whose temperature and chemical potential  
is chosen self-consistently so that there is no average flux of 
particles and energy between the probe itself and the system under examination.
The advantage of such approach lies in its simplicity and independence
of microscopic details of inelastic processes.   
Probe terminals have been widely used in the literature
and proved to be useful to unveil nontrivial aspects of 
phase-breaking
processes \cite{datta}, heat transport and rectification 
\cite{visscher,lebowitz,dhar2007,dhar,lebowitz2008,pereira,segal},
and thermoelectric transport
 \cite{jacquet2009,imry2010,buttiker2011,saito2011,sanchez2011,
imry2012,buttiker2012}.

\begin{figure}[!htb]
\centering
\includegraphics[width=5cm]{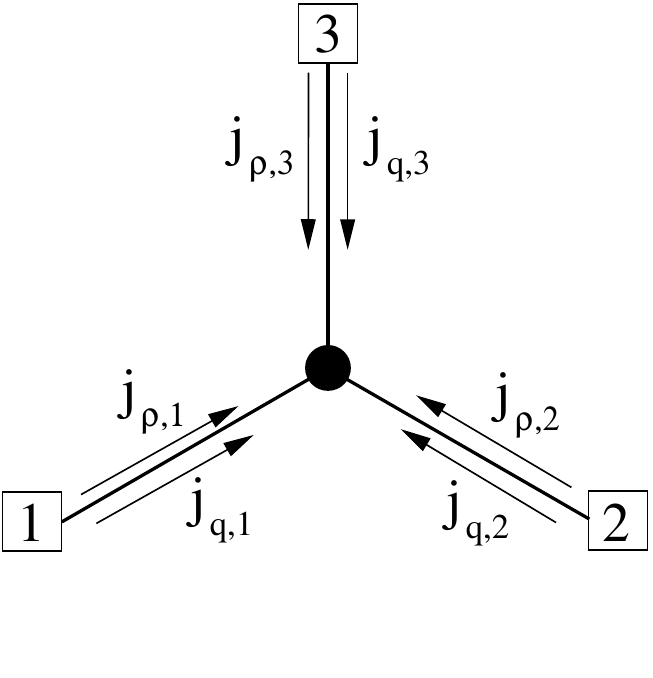}
\caption{Schematic drawing of the three terminal transport model,
with $j_{\rho,i}$ and $j_{q,i}$ denoting 
the particle and heat currents from terminal $i$.}
\label{fig:mercedes}
\end{figure}

In this Letter we propose and study a simple multi-terminal classical 
transport model, where
the wires connected to the system are one-dimensional, i.e. they have 
just one momentum state per each value of the energy, and where the
\emph{deterministic} dynamics of the system, 
at fixed energy, is simply a permutation 
(rewiring) among the terminals. 
In particular, we will focus on three-terminal  
models, depicted in Fig.~\ref{fig:mercedes}, where the second terminal is 
considered as a probe. 
We discuss the coupled heat and particle  
deterministic transport between the remaining wires within the linear 
response limit. In this limit the transport between the first and the 
third wire is described by the reduced Onsager matrix. As our energy-dependent 
rewiring is in general not self-inverse we are particularly interested in the 
precise conditions for the reduced Onsager matrix to be asymmetric and 
discuss how to maximize the asymmetry.

\par
{\em Classical coupled transport formalism.-}
Let us consider a generic $N$-terminal, non-interacting 
classical transport model.  
Each wire labelled by $i$ is connected to a thermo-chemical bath at reciprocal temperature 
$\beta_i=\beta + \delta \beta_i$ and chemical potential 
$\mu_i = \mu +\delta \mu_i$ ($i=1,...,N$), 
where $\beta$ and $\mu$ are reference 
reciprocal temperature and chemical potential, respectively. 
The $\delta \beta_i$ and $\delta \mu_i$ are considered as gradient 
fields w.r.t. reference values. Such gradients are in the linear 
response regime small in magnitude.
The particles are effused from the wires into the junction, 
with the injections rates  
\beq
  \gamma_i = \gamma \frac{\beta}{\beta_i} e^{\beta_i \mu_i - \beta \mu},
\eeq
where $\gamma$ denotes the injection rate at the reference values
$\beta$ and $\mu$. Without loss of generality, we set 
$\gamma=\beta=1$, and $\mu=0$.
At inverse temperature $\beta$, the particles energies 
are distributed according to Boltzmann's formula 
$p(E) = \beta e^{-\beta E}$. 
In the stationary, linear response regime, we have in the $i$th wire a particle 
current $j_{\rho,i}$ and a heat current $j_{q,i}$ proportional 
to gradients $\delta \beta_j$ and 
$-\beta\delta \mu_j=-\delta\mu_j$ \cite{degroot}. 
To state this more precisely, we  introduce the 2-vector of currents 
$J_i=(j_{q,i}, j_{\rho,i})$ in the $i$th terminal, and the 
2-vector of gradient fields   
$X_j=(\delta \beta_j, -\delta \mu_j)$ in the $j$th terminal,
and $2\times 2$ blocks of the Onsager matrix 
\beq
  L_{i,j} = \left [
  \begin{array}{cc} K_{i,j} & Q_{i,j} \\ Q_{i,j} & T_{i,j}\end{array}
  \right] \>,
\label{eq:blockonsager}
\eeq
connecting the two of them:
\beq
  J_i =  \sum_{j=1}^N L_{i,j} X_j \>.
\eeq
The matrix elements of $L_{i,j}$ are defined by 
\beq
  (T_{i,j},Q_{i,j},K_{i,j}) = 
  \int_{\bR_+} \dd E\, e^{-E} \tilde\tau_{i,j}(E)(1,E,E^2) \>,
\eeq
with $\tilde \tau_{i,j} \equiv \delta_{i,j}-\tau_{i\gets j}$,
and the \emph{on-shell transmission functions} $\tau_{i\gets j}$
satisfying the probability conservation $\sum_i \tau_{i\gets j}(E) = 1$
and the sum rule $\sum_i \tau_{i\gets j}(E)=
\sum_j \tau_{i\gets j}(E)$, ensuring that the currents 
vanish when all the potentials and temperatures are equal. 
The conservation of the net currents translates to a 2-vector 
condition $\sum_{i=1}^N J_i = 0$. 
We note that due to conservative and non-interacting nature of our model, 
the Onsager matrix is always, as explicitly written in 
Eq.~(\ref{eq:blockonsager}), block symmetric. 
This property is a consequence of conservation of total probability
and would translate also in any quantum extension of our model, 
being in that case a consequence of the unitarity of 
the S-matrix \cite{buttiker}.

{\em The railway switch model.-}  We consider a \emph{deterministic} 
transport model in which the outgoing wire for a particle is uniquely
determined by the incoming wire and the particle's energy. 
Such model is not bound to be symmetric with respect to 
time reversal, and therefore allows us to systematically 
study the effects of time reversal symmetry breaking.
The model is fully specified by transmission functions
$\tau_{i\gets j} (E)$ for which only the values zero and one
are allowed: $\tau_{i\gets j} (E) = 1$
if particles injected from terminal $j$ with energy $E$ go 
to terminal $i$, $\tau_{i\gets j} (E) = 0$ if such particles go
to a terminal different from $i$.
The above deterministic on-shell transmission 
functions can be described using permutation  
matrices $P_k$ for $k \in I\equiv\{1,\ldots,N!\}$ corresponding 
to the permutation group ${\cal S}_N$ as
\beq
  \tau_{i\gets j}(E) = [P_{\psi(E)}]_{i,j}\qquad i,j=1,...,N\>,
  \label{eq:tau_ij}
\eeq
where $\psi(E): \bR \to I$ is a piecewise constant function controlling 
the (energy-dependent) switching between the permutations. 
The function $\psi(E)$ for $n$ permutation switches is completely specified 
in terms of a sequence of $n+1$ integers 
$\pi=(p_0,p_1,\ldots,p_n)$, $p_i\in I$, and a sequence of 
$n$ threshold energies $\epsilon=(E_1,\dots,E_n)$, at which 
the switches occur:
\beq
  \psi(E) = p_i \in I \qquad \textrm{for}\quad E \in [E_i,E_{i+1}] \>,
\eeq
with $i = 0, \ldots, n$, $E_0 \equiv 0$ and $E_{n+1} \equiv \infty$. 
That is to say, at energy $E_i$ ($i=1,...,n$) we switch from 
permutation $P_{p_{i-1}}$ to $P_{p_i}$.
A {\em realization} of the model is defined by a pair $(\pi,\epsilon)$.

From now on we focus on the ($N=3$)-terminal case
(Fig.~\ref{fig:mercedes}), so that the 
permutation matrices $P_k \in \{0,1\}^{3\times3}$ read 
\beqaS
  P_1 = \left[ 
  \begin{array} {ccc} 
  1 & 0 & 0 \\
  0 & 1 & 0 \\
  0 & 0 & 1 
  \end{array}
  \right]
  \begin{minipage}[l]{1cm}
  \includegraphics[width=1cm]{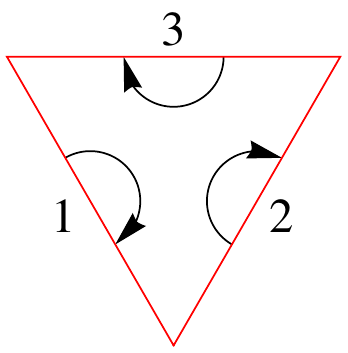}
  \end{minipage},\>
  P_2 = \left [
  \begin{array} {ccc} 
  1 & 0 & 0 \\
  0 & 0 & 1 \\
  0 & 1 & 0 
  \end{array}
  \right]
  \begin{minipage}[l]{1cm}
  \includegraphics[width=1cm]{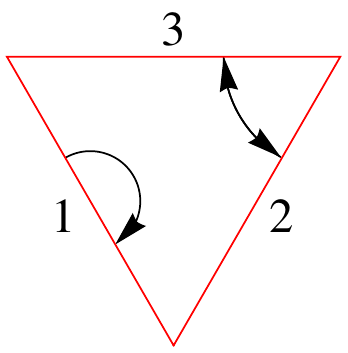}
  \end{minipage},
  \\
  P_3 = \left[ 
  \begin{array} {ccc} 
  0 & 1 & 0 \\
  1 & 0 & 0\\
  0 & 0 & 1 
  \end{array}
  \right]
  \begin{minipage}[l]{1cm}
  \includegraphics[width=1cm]{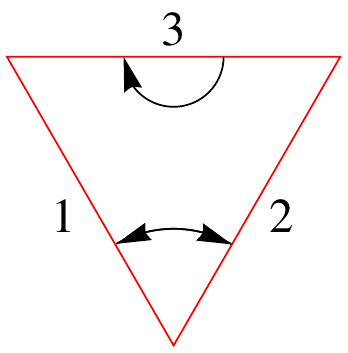}
  \end{minipage},\>
  P_4 = \left[ 
  \begin{array} {ccc} 
  0 & 1 & 0 \\
  0 & 0 & 1 \\
  1 & 0 & 0 
  \end{array}
  \right]
  \begin{minipage}[l]{1cm}
  \includegraphics[width=1cm]{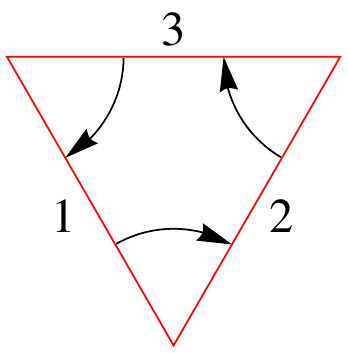}
  \end{minipage},\\
  P_5 = \left [
  \begin{array} {ccc} 
  0 & 0 & 1 \\
  1 & 0 & 0 \\
  0 & 1 & 0 
  \end{array}
  \right]
  \begin{minipage}[l]{1cm}  
  \includegraphics[width=1cm]{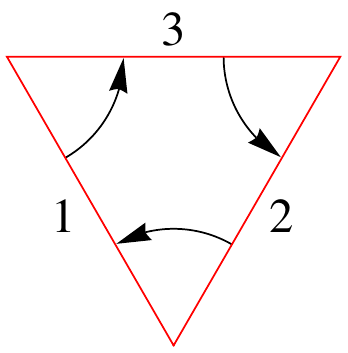}
  \end{minipage},\>
  P_6 = \left [
  \begin{array} {ccc} 
  0 & 0 & 1 \\
  0 & 1 & 0 \\
  1 & 0 & 0 
  \end{array}
  \right]
  \begin{minipage}[l]{1cm}  
  \includegraphics[width=1cm]{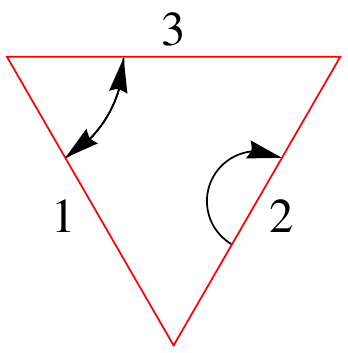}
   \end{minipage}.
\eeqaS
To the right of each permutation we show the rewiring described by 
the matrix. Following the figures we see that each permutation has a 
very simple interpretation, e.g., $P_1$ represents the reflection of 
the incoming particles to the original terminals, 
in $P_2$ particles from terminal $1$ are reflected, particles
from terminal $2$ go to terminal $3$ and particles from 
terminal $3$ go to terminal $2$, etc.

Our model has interesting symmetries.
By introducing the super-operator of time inversion 
$\hat\cT X = X^T$ (with $X^T$ the transpose of $X$), we can see that
\beq
  \hat\cT P_k = P_k\quad\textrm{for}\quad k\in\{1,2,3,6\}
  \quad\textrm{and}\quad
  \hat\cT P_4 = P_5 \>.
\eeq
This implies that $P_4$ and $P_5$ are the only time-asymmetric permutations, 
with $P_4 = P_5^{-1}$ and $P_5 = P_4^{-1}$, 
while for all other cases we have $P_k=P_k^{-1}$. 
By considering the second terminal to act as a probe, the model is 
invariant on swapping the first and the third terminal.
These operation can be performed by the swap super-operator 
$\hat\cS X = P_6 X P_6$ acting as
\beq
  \hat\cS P_1 = P_1\>,\quad
  \hat\cS P_2 = P_3\>,\quad
  \hat\cS P_4 = P_5\>,\quad
  \hat\cS P_6 = P_6\>.
\eeq

\emph{Thermoelectric transport with a probe terminal.-} 
To illustrate the railway switch model
in a concrete example, we discuss 
thermoelectric (or thermochemical) transport,
when the second terminal acts as a probe, i.e., $J_2=0$.
We choose to measure gradients w.r.t. the third wire by setting $X_3=0$.
We end up relating the remaining gradient fields
\beq
  X_2 = - L_{22}^{-1} L_{21} X_1,
\eeq
and connecting the currents between the first and the third wire,
\beq
  J_1 = L_{\rm red} X_1,
\eeq
in terms of a {\em reduced Onsager matrix} $L_{\rm red}$, defined as
\beq
  L_{\rm red} = L_{11} - L_{12} L_{22}^{-1} L_{21}  
  = \left [ 
  \begin{array}{cc}
  l_{11} & l_{12} \cr
  l_{21} & l_{22}
  \end{array}
  \right] \>.
\eeq
The reduced Onsager matrix admits a nice analytic description of matrix 
element $l_{i,j}$. Let us first introduce the determinant $D=\det L_{22}$ 
of the Onsager matrix $L_{22}$ for the transmission from the probe, written as
\beq
 D = \frac{1}{2} \int_{\bR_+ \times \bR_+} 
 \left[\prod_{i=1}^2 e^{-E_i} \tilde\tau_{22}(E_i) \dd E_i \right]
(E_1 - E_2)^2
\eeq
and define the integration measure 
\beq
  \dd \mu = \left[\prod_{i=1}^3 e^{-E_i} \dd E_i \right]
  \tilde\tau_{21}(E_1)\tilde\tau_{12}(E_2)\tilde\tau_{22}(E_3)
\eeq
over the domain $(E_1,E_2,E_3)\in \cD = R_{+}^3$. (To simplify notation we omit the arguments $(E_1,E_2,E_3)$ in the measure $\dd \mu$.) 
Then we can write the matrix elements of $L_{\rm red}$:
\beqa
 l_{11} &=& K_{11} -\frac{1}{D}\int_\cD \dd \mu\,  E_1 E_2 (E_1-E_3)(E_2-E_3), \\
 l_{22} &=& T_{11} -\frac{1}{D}\int_\cD \dd \mu\, (E_1-E_3)(E_2-E_3), \\
 l_{12} &=& Q_{11} -\frac{1}{D}\int_\cD \dd \mu\, f(E_1,E_2,E_3), \\
 l_{21} &=& Q_{11} -\frac{1}{D}\int_\cD \dd \mu\, f(E_2,E_1,E_3), 
\eeqa 
with $f(E_1,E_2,E_3) \equiv E_2 E_3^2 + E_1 E_2^2-(E_1 + E_2)E_2 E_3$. 
It is convenient to introduce the quantities 
\beqa
2 \mean{f} &:=& f(E_1,E_2,E_3)+f(E_2,E_1,E_3) \nonumber \\
           &=& (E_1+E_2)(E_1-E_3)(E_2-E_3)\>, \\
2 \Delta f &:=& f(E_1,E_2,E_3)-f(E_2,E_1,E_3) \nonumber \\
           &=& \sum_{i,j,k} \epsilon_{i,j,k} E_j E_k^2\>,
\eeqa
with $\epsilon_{i,j,k}$ being the Levi-Civita totally asymmetric tensor. 
Note that $L_{\rm red}$ can be decomposed into the sum of $L_{11}$ and 
a matrix defining the communication between the wires. 

In this work we are mainly interested in the asymmetry measure $x$ of 
the reduced Onsager matrix, defined as
\beq
  x = \frac{l_{21}}{l_{12}} \>.
  \label{eq:asymmX}
\eeq
This parameter has been discussed for quantum dots with broken 
time-reversal symmetry \cite{saito2011,footnote}. 
If the transport is time-reversal symmetric, then 
$\int \dd\mu \Delta f=0$ and consequently $x=1$.
Large asymmetries are desirable since they could in 
principle lead to high thermoelectric efficiencies. 
Indeed, by considering the reduced 
system as a heat engine with steady state power generation 
$P=-j_{\rho, 1} \delta \mu_1$, dissipated heat 
current $Q=j_{q, 1}$, and efficiency $\eta=P/Q$,
the maximum efficiency reads  
\cite{benenti2011} 
\beq  \eta_{\rm max}  = \eta_{\rm C} x \frac{\sqrt{1+y}-1}{\sqrt{1+y}+1} \>,
\label{eq:efficiency}
\eeq
where $\eta_{\rm C}=|\delta\beta_1|$ (for $\beta=1$) is the 
Carnot efficiency and $y = \frac{l_{12}l_{21}}{\det L_{\rm red}}$ is the figure of merit.
Note that $x$ acts as a multiplier to the efficiency and so its maximization 
(at fixed $y$) is desirable for increasing the efficiency. 
The thermodynamic bounds on $x$, $y$, and $\eta$
are discussed in Ref. \cite{benenti2011}.

In the following we discuss the asymmetry measure $x$ and 
the figure of merit $y$ of the reduced Onsager matrix 
in different realizations of the railway switch model model. 
At fixed threshold energies $\epsilon$, by considering the sequence of 
inverted permutations we obtain the same figure of merit and reciprocal 
measure of asymmetry~\cite{footnote2}:
\beq
  y ( \hat \cT \pi) = y (\pi)\>,\qquad  x ( \hat \cT \pi ) =  x (\pi)^{-1}\>,
\label{eq:reciprocal}
\eeq
while exchanging the first and the third wire does not affect 
the two quantities:
\beq
y ( \hat \cS  \pi) = y (\pi)\>,\qquad   x (  \hat \cS \pi )= x (\pi) \>.
\eeq
If a realization is symmetric w.r.t. to 
$ \pi\to \hat \cT \pi$, then is has a 
symmetric Onsager matrix and so $x=1$. We observed numerically  
that the Onsager matrix of our model has all elements positive, 
$l_{i,j}>0$, so that the asymmetry measure $x$  is always non-negative, $x\ge 0$. 

\begin{figure}[!htb]
\centering
\includegraphics[width=8cm]{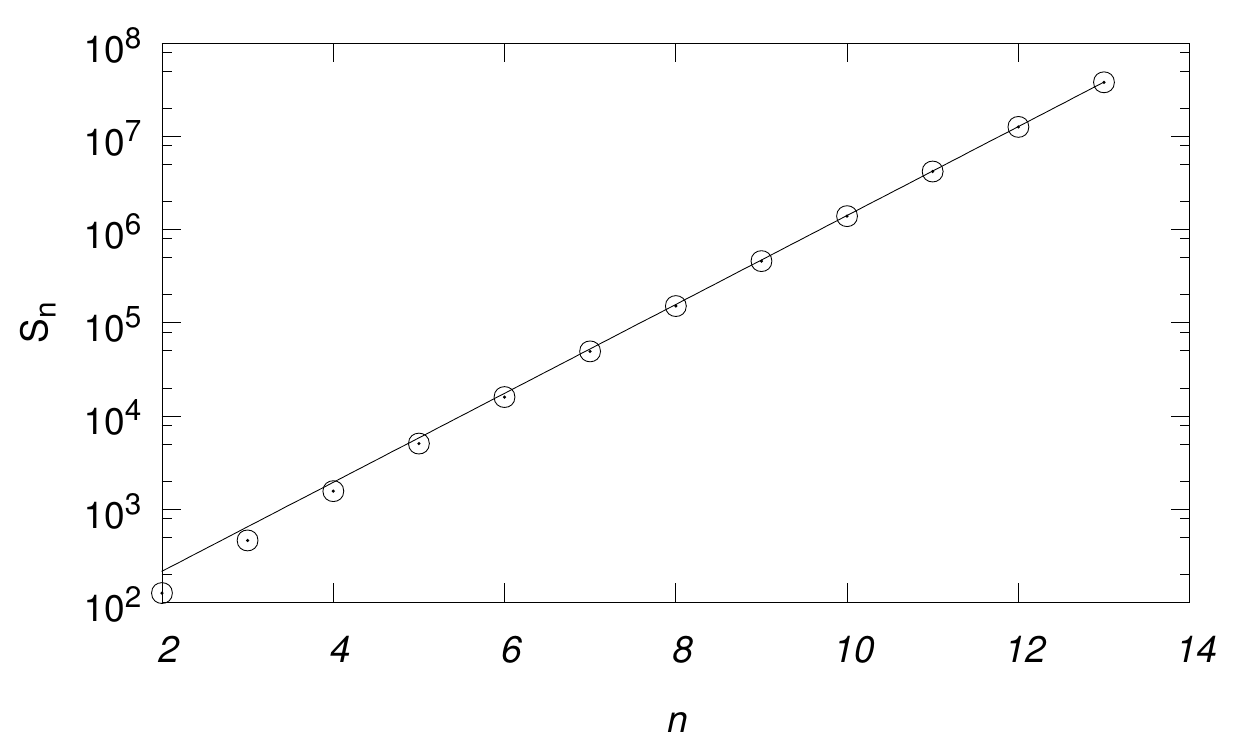}
\caption{The number of cases $S_n$ with symmetric Onsager 
matrix as a function of the number $n$ of switches and best fitted asymptotic dependence $24\cdot 3^n$.}
\label{fig:stat}
\end{figure}

The number of all $n$-switch cases is $N_n=6\cdot 5^n$. 
Let $S_n$ denote the number of cases for which we have $x=1$. 
As shown in Fig.~\ref{fig:stat}, 
\beq
  S_n \sim C\, \alpha^n\quad\textrm{as}\quad n\to\infty\>,
\eeq 
with $\alpha\approx 3$ and $C\approx 24$. We see that $S_n$ is asymptotically approximately six times larger than the 
number of cases composed of only symmetric permutations, 
equal to $\tilde S_n = 4\cdot 3^{n}$. 
The number of cases $A_n=N_n-S_n$ with an asymmetric Onsager matrix 
increases exponentially, as ${\cal O}(5^n)$. Hence, most of 
the cases lead to $x\ne 1$.

In the following, we discuss the properties of our model for different number 
$n$ of switches, thus increasing with $n$ the complexity of the model
in a controllable manner.
All one-switch ($n=1$) cases have $x= 1$. The asymmetry 
$x\not = 1$ is possible only in the cases with two or more switches. 
In the two-switch ($n=2$) cases only non-repeated combinations of 
permutations $P_k$ with indices in $k \in \{2,3,4,5\}$ produce 
asymmetry, but the asymmetry parameter $x$ is 
always limited to a finite interval, namely it is 
always strictly larger than zero and finite.
In the three-switch ($n=3$) cases the asymmetry parameter 
can, for specific sequences of permutations, 
become arbitrarily large.
This fact is possible for the following sequences of permutations:
\beqa
 V = \{ (2, 3, 1, 4),\, (3, 2, 1, 5),\, (4, 2, 1, 3),\, \\\nonumber 
(4, 2, 1, 5),\, (5, 3, 1, 2),\, (5, 3, 1, 4) \}\>.
\eeqa
The obtained set is invariant w.r.t. 
$\hat{\cal S}$, swapping the first and the third channels. 
Therefore we can limit ourselves to consider a desymmetrized set
\beq
 \widetilde V = \{ (2, 3, 1, 4),\, (4, 2, 1, 5),\, (4, 2, 1, 3) \}\>.
  \label{eq:p3max}
\eeq
Equally interesting are the cases in which $x$ limits to 0, which
are obtained by time-inverting 
($\pi\to \hat \cT \pi$)
the cases in $V$. In all 
these cases we can tend to the maximal asymmetry, provided the 
switch energy thresholds are chosen with 
certain asymptotic scalings. By expressing the threshold energies 
$E_{i+1} = E_i + \Delta E_i$
in terms of the gaps $\Delta E_i>0$, for the cases of $V$, a local 
maximum $x_{\rm max}$ of $x$ at fixed 
$\Delta E_2$ follows a curve scaling in the limit 
$\Delta E_2\to\infty$ as
\beq  
  E_0 \asymp e^{-\alpha \Delta E_2}\>,\quad 
  \Delta E_1 \asymp e^{-\beta \Delta E_2}, \>
\eeq
and asymmetry $x$ diverges along this curve according to
\beq
  x_{\rm max} \asymp e^{(\alpha -\beta)  \Delta E_2} \>.
\eeq  
For the first two cases in $\widetilde V$ we find  $(\alpha,\beta) = 
(0.488, 0.316)$, for the last one  $(\alpha,\beta) \doteq (0.842, 0.536)$. 
We note that three-switch cases of $V$ are obtained from two-switch asymmetric 
cases by inserting permutation 
$P_1$ into the third position. We maximize the asymmetry by increasing 
the energy interval controlled by $P_1$ permutation and decrease all other
intervals. 

\begin{figure}[!htb]
\centering
\includegraphics[width=7.5cm]{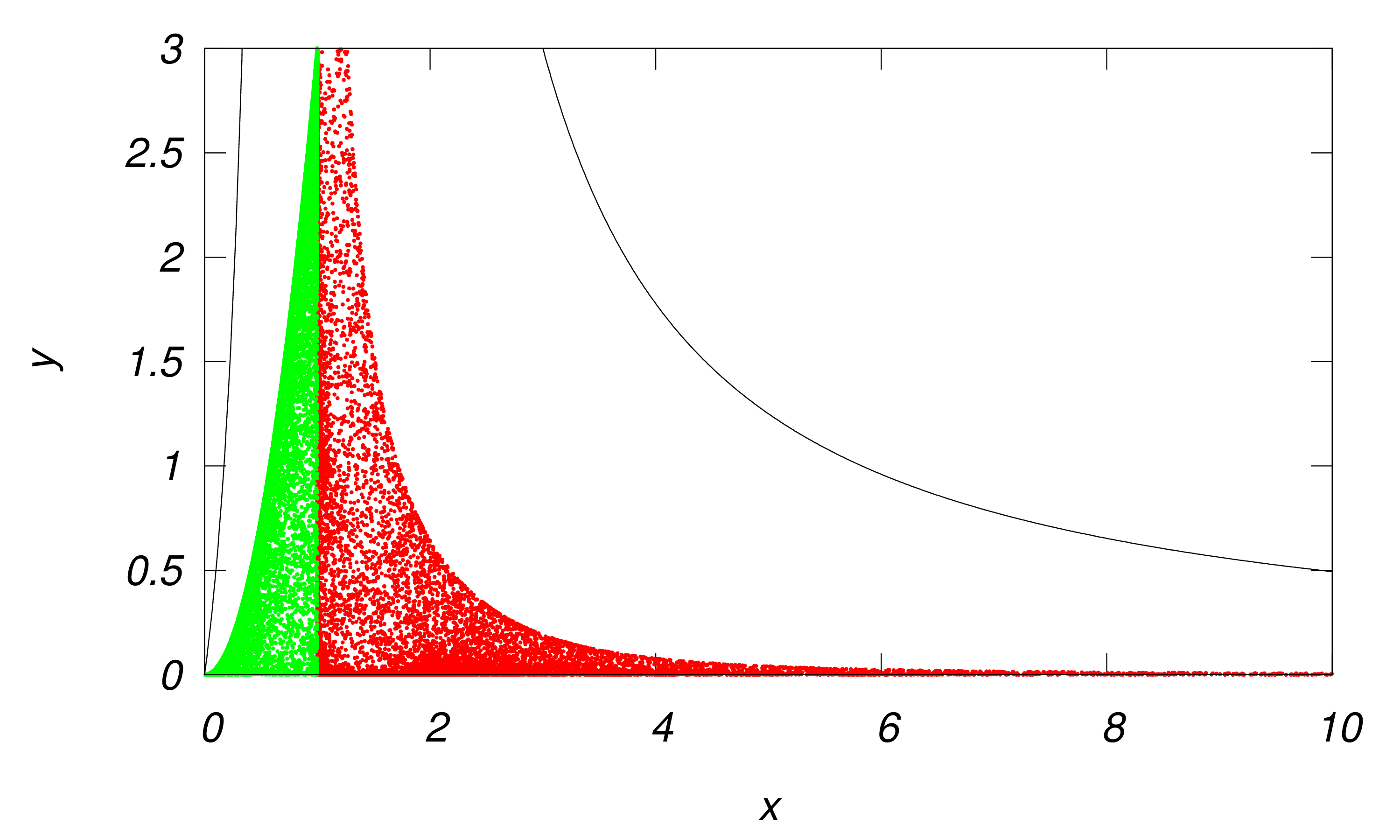}
\caption{The relation between $x$ and $y$ for random choices of threshold 
energies in the case $(2, 3, 1, 4)$ (red dots) and 
in the time inverted case of $(4, 2, 1, 3)$, that is, in the case 
$(5, 2, 1, 3)$ (green dots). 
The black curve gives the analytic boundary $y=4x/(x - 1)^2$.}
\label{fig:xy3}
\end{figure}

In Fig.~\ref{fig:xy3} we show how the figure of merit $y$ is 
related to the asymmetry measure $x$ in two examples, one from
$\pi\in \widetilde V$ and the other corresponding to its time-inversion 
$\pi\to \hat \cT \pi$.
We see that increasing the asymmetry $x$ results on average in decreasing 
the figure of merit $y$, in such a way that we do not improve 
the efficiency $\eta$. Note that, for $x\ge 1$, we are far from 
the Carnot efficiency, which is achieved on the curve 
$h(x)=4x/(x - 1)^2$~\cite{benenti2011} (black curve in Fig.~\ref{fig:xy3}). 
Further increasing the number of switches the diversity of cases increases 
beyond the ability of a detailed case-by-case study. It remains an open
problem whether one could with our model and for a large number
of switches, approach the Carnot 
efficiency at asymmetries $x\gg 1$. 

{\em Generalizations - } By replacing permutation matrices 
$P_i\in {\cal S}_3$ with matrices 
corresponding to the permutation group of arbitrary degree $N$, our model 
describes deterministic transport in an $N$-terminal junction of one-dimensional 
wires. Seeking optimized transport  in our model therefore corresponds 
to discrete optimization problems on permutation groups. As a principal, but 
completely impractical generalization we note that we can phrase any 
deterministic scattering dynamics in terms of our switch model.
Finally, our model could be easily extended to investigate 
nonlinear regimes where breaking of time-reversibility
has nontrivial effects on the transport \cite{buttiker2004},
and reformulated in a quantum mechanical context.

{\em Conclusions -} We have presented a minimalistic classical 
finite-state deterministic transport model which allows to systematically 
study the questions of coupled particle and heat transport and 
thermoelectric/thermochemical efficiency. We have analyzed in particular 
the three-terminal model, where one terminal serves as the probe, 
and analyzed in detail the conditions under which one can maximize the 
asymmetry of the reduced Onsager matrix in relation to time-reversal breaking 
in the model. We expect that our model may serve as a simple benchmark 
for mesoscopic coupled transport studies.
M.H. and T.P acknowledge supported by the grant
  P1-0044 of Slovenian Research Agency, and  G.B. and G.C. by the MIUR-PRIN 2008 and
by Regione Lombardia.


\begin{thebibliography}{99}
\bibitem{footnote}
An alternative measure of asymmetry can be defined as
\beq  
z = \frac{l_{21}-l_{12}}{l_{12}+l_{21}} = 
\frac{\int \dd \mu\,\Delta f} {Q_{11} D - \int \dd \mu\, \mean{f}} \>,  
\label{eq:asymmZ}
\eeq
which is related to $x$ (\ref{eq:asymmX}) via $ x = (1+z)/(1-z)$. 
In the time-reversal symmetric case $l_{12}=l_{21}$ and therefore $z=0$.
In this paper we prefer to focus on parameter $x$ rather than $z$, 
as it directly appears in formula (\ref{eq:efficiency}) for 
maximum thermoelectric efficiency.
\end{thebibliography}
\end{document}